\documentclass[aps,twocolumn,prl]{revtex4}

\usepackage{graphicx}
\usepackage{bm,color}

\usepackage{graphicx}
\usepackage{bm,color}
\usepackage{multirow}

\newcommand{\be}{\begin{eqnarray}}
\newcommand{\ee}{\end{eqnarray}}

\renewcommand{\theequation}{\arabic{equation}}

\begin{document}

\title{Many--body--localization induced protection of symmetry--protected--topological order in an XXZ spin model}
\date{\today}

\author{Yoshihito Kuno}
\affiliation{Department of Physics, Graduate School of Science, Kyoto University, Kyoto 606-8502, Japan}

\begin{abstract}
There is a counter-intuitive expectation proposed by Huse {\it et al} [Phys. Rev. B {\bf 88}, 014206 (2013)], 
and Chandran {\it et al} [Phys. Rev. B 89,144201 (2014)]: 
Localization protects quantum order of groundstate even in high excited eigenstates.
In this work, we numerically investigate the localization protection for symmetry--protected--topological (SPT) order 
by considering a modified XXZ spin model, related to an interacting Su--Schrieffer--Heeger (SSH) model. 
We systematically study how Many--body--localization (MBL) protects SPT for different disorder types and different forms of interactions. 
A certain disorder leads the clear degeneracy of the low--lying entanglement spectrum of each excited many--body eigenstates. 
This fact indicates that existence of edge modes, which is a hallmark of the SPT order, is protected by the MBL in excited many--body eigenstates.
In addition, we also report how the MBL protected edge modes in excited many--body eigenstates fade away for a diagonal type disorder. 
\end{abstract}


\maketitle
\textit{Introduction.---}
Many--body--localization (MBL) is one of the most active topics in condensed matter physics. 
Theoretically, beyond the classical studies of the Anderson localization \cite{Anderson, Lagendijk}, 
rich physical phenomena have been expected so far \cite{Nandkishore,Basko,Abanin1,Alet,Abanin2,Oganesyan1,Iyer1}.
Experimentally, recent cold atom experiments have claimed to capture the ergodicity breaking dynamics of the MBL \cite{Schreiber,Choi,Lukin,Rispoli}.
On the other hand, topological state of matter is also one of the important topics in condensed matter physics. 
Various artificial topological states of matter have been realized in recent experimental systems \cite{Cooper,Ozawa} 
and theoretical predictions have been simulated.
In this work, we study the interplay of MBL and topological state. 
Recently, Huse {\it et al.} \cite{Huse} and Chandran {\it et al.} \cite{Chandran} proposed a unique conjecture about the Anderson localization and MBL:
A groundstate quantum order is protected even in excited many--body eigenstates, i.e., the MBL protects the groundstate quantum order.  
Shortly after proposing the conjecture, 
Kjall {\it et al.} \cite{Kjall} and Pekker {\it et al.} \cite{Pekker} have tested the MBL induced protection of ferromagnetic ${\bf Z}_2$-symmetry breaking order 
by considering a one-dimensional (1D) quantum random Ising model. 
Bahri {\it et al.} \cite{Bahri} have also considered a 1D disordered cluster spin model as a test model 
and confirmed the MBL induced protection of the symmetry--protected--topological phase (SPT)\cite{SPT1,SPT2}. 
This remarkable result indicates that the SPT order under the MBL could in principle exist out of equilibrium 
and hence, cooling to detect the SPT order in real experiments is unnecessary\cite{Chandran,Parameswaran}.
Furthermore, recently, for the 1D disordered cluster spin model, 
the detailed phase diagram including co-existence of the MBL and SPT order has been investigated \cite{Dekker}.
In addition, the relationship between the MBL and SPT phase 
for a time-reversal symmetric model has been extensively discussed from the perspective of the matrix product state \cite{Wahl,Wahl2}.
However, the following issues require further investigation: 
(I) Does the conjecture stated in \cite{Huse,Chandran} hold universally for various condensed matter models? 
(II) To what extent in the whole spectrum does the mechanism of the MBL protection of quantum order work? 
(III) How strong does the quantum order of groundstate remain in many--body excited states? 
In this work, we consider a modified XXZ spin model relating to the Su-Schrieffer-Heeger (SSH) model \cite{SSH}, 
which is one of the simple models exhibiting 1D non-trivial topological insulating phase corresponding to the SPT order \cite{Asboth},
 and numerically investigate the conjecture in \cite{Huse,Chandran}. 

In addition, whether the SPT phase in an SSH type model or the AKLT model is protected by the MBL remains an open question \cite{Chandran}. 
Recently, a study \cite{Vasseur} indicated that 
an XXZ model with a particle-hole symmetric disorder and an SSH-type coupling exhibits no protection of the SPT phase by the MBL. 
Here, for various conditions of disorder and interactions, 
we numerically evaluate the protection of the SPT phase and also how effective the arguments in \cite{Vasseur} are.

\begin{figure}[t]
\centering
\includegraphics[width=6.5cm]{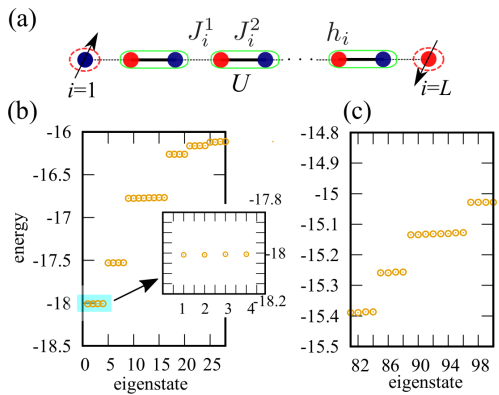}
\caption{(a) Modified XXZ spin model with open boundary condition. 
The red-dotted circles on either ends represent the edge sites of the chain. 
In SPT phase ($J_2>J_1$), the edge mode appears as free spin--1/2 degree of freedom, 
corresponding to zero energy excitation (in thermodynamic limit). 
(b) Lowest to $20$-th energy spectrum for $L=18$ system.
(c) $81$-th to $100$-th energy spectrum with the same parameters as in (b).}
\label{Setup}
\end{figure}

\textit{Modified XXZ spin model and interacting SSH model.---}
We consider a 1D XXZ spin model with dimerized coupling: 
\begin{eqnarray}
H_{XXZ}&=&\sum_{i}\biggl[J_{i}(S^{+}_{i}S^{-}_{i+1}+S^{-}_{i}S^{+}_{i+1})+U_{i} S^{z}_{i}S^{z}_{i+1}+h_{i}S^{z}_{i}\biggr],\nonumber
\label{XXZspin}
\end{eqnarray}
where $S^{+(-)}_{i}$ and $S^{z}_{i}$ are spin--1/2 raising (lowering) operator and spin--1/2 $z$-component operator, 
$i$ denotes lattice site taking $i=1,\cdots, L$, where $L$ is a system size. The model parameters are shown in Fig~\ref{Setup} (a), 
for $i\in odd$, $J_{i}\to J^{o}_{i}=J_{1}+\delta J^{o}_{i}$, for $i\in even$, $J_{i}\to J^{e}_{i}=J_{2}+\delta J^{e}_{i}$, 
where $\delta J^{o(e)}_{i}$ is a uniform disorder, $[-\delta J_{1(2)} ; \delta J_{1(2)}]$, and $h_{i}$ 
is also a uniform disorder, $[-\delta h ; \delta h]$. 
$U_{i}$ is regarded as an interaction in a fermion picture and in this work acts only on the $i\in even$ links with a strength $U$. 
This interaction pattern $U_{i}$ may be implemented by considering a combination of a double well and a spin dependent optical lattice \cite{SSH_ours}.
Using the Jordan-Wigner transformation, the XXZ model of $H_{XXZ}$ can be mapped into an interacting SSH model \cite{Sirker,SSH_ours} with disorder
\begin{eqnarray}
H_{SSH}&=&\sum_{i}\biggl[J_{i}(f^{\dagger}_{i}f_{i+1}+\mbox{h.c.})\nonumber\\
&+&U_{i} (n_{i}-1/2)(n_{i+1}-1/2)+h_{i}n_{i}\biggr],\nonumber
\label{SSHmodel}
\end{eqnarray}
where $f^{(\dagger)}_{i}$ is an annihilation (creation) spinless fermion and $n_{i}=f^{\dagger}_{i}f_{i}$ is a number operator.
Based on this model, let us note the topological properties. In the clean limit and $U=0$ non-interacting case, 
the model belongs to the BDI class in the topological classification \cite{Schnyder,Kitaev}. 
At half-filling, the SSH model exhibits a non-trivial topological phase.
The non-trivial topological phase is known to be the SPT phase protected by the chiral symmetry. 
The SPT phase remains even for a finite $U_{i}$ \cite{Sirker,SSH_ours,Yahyavi,Hetenyi,Weber}. 
The model has time-reversal and particle-hole symmetry \cite{Vasseur}, 
then the interacting SSH model is invariant under the chiral (sub-lattice) transformation, $f^{\dagger}_{i}\to (-1)^{i}f_{i}$, $f_{i}\to (-1)^{i}f^{\dagger}_{i}$ \cite{Manmana}. 
In a disordered case, since $\delta J^{o}_{i}$ and $\delta J^{e}_{i}$ disorders do not break the chiral symmetry, 
the groundstate SPT phase is expected to be fairly robust to them.  
In contrast, if the disorder $h_{i}$ is diagonal so that the chiral symmetry is broken, the groundstate SPT becomes fragile. 
Since the XXZ model of $H_{XXZ}$ inherits the topological properties of the SSH model of $H_{SSH}$ \cite{Estarellas}, 
we numerically treat the XXZ model of $H_{XXZ}$.
In what follows, for all numerical simulations, we fix $J_{1}=0.1$, $J_{2}=1$, and $\delta h=0.01$\cite{degeneE} except for Fig.~\ref{Fig4}.

\begin{figure*}[t]
\centering
\includegraphics[width=18cm]{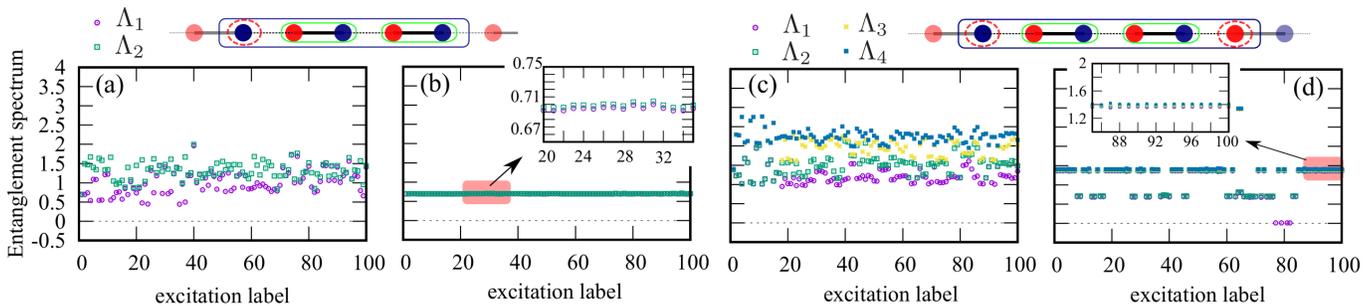}
\caption{Low-lying entanglement spectrum for low energy excited eigenstates. 
(a) Lowest and next-lowest entanglement spectrum in $L=14$ clean system with $U=0$, 
where the partial system cut has one free 1/2 spin on the left edge. 
(b) Lowest and next-lowest entanglement spectrum in $L=14$ disordered system with $U=2$ and $\delta J_{2}=3$. 
(c) Lowest to fourth entanglement spectrum in $L=16$ clean system with $U=0$, 
where the partial system cut has two free 1/2 spins on both left and right edges in the SPT order. 
(d) Lowest and fourth entanglement spectrum in $L=16$ disordered system with $U=2$ and $\delta J_{2}=3$.}
\label{ES_degenracy}
\end{figure*}
\textit{Phenomenology: Localization protects SPT order.---}
The studies by Huse et al. \cite{Huse} and Chandran et al. \cite{Chandran} state that 
a quantum order of groundstate in an MBL system with weak interactions is protected in low energy excited states.
The low-energy excitations above the groundstate are expected to be localized due to a disorder. 
Since very weak interacting (almost non-interacting) excitation particles are in Anderson localization, 
the excitations are suppressed to extend. 
As a concrete example, for 1D quantum random Ising model the groundstate is ferromagnetic. The excitation is a kink (domain wall) excitation. 
The excitation kinks are localized owing to disorder, and therefore, the ferromagnetic domains are stable. 
This leads to a spin--glass order even in high excited eigenstates \cite{Kjall,Pekker}. 

The above phenomenon may apply to the model of $H_{XXZ}$. 
For $L= even$ clean system with $J_{1}\ll J_{2}$ and open boundary, 
the groundstate of the system exhibits a valence--bond--solid order in the bulk, 
where the singlet state is formed on $J^{e}_{i}$--links. 
The groundstate is in the SPT phase because on the left and right edge sites a free spin--1/2 degree of the freedom exists (See Fig.~\ref{Setup} (a)) 
corresponding to zero--energy edge mode \cite{edge_mode}. 
The edge modes induce fourfold degeneracy of the groundstate. 
Here, let us consider $2J_{2}=U$ case and a single excitation on a single $J^{e}_{i}$--link. 
The singlet groundstate on the link excites to three triplet states. 
The excitation may be regarded as a triplon as in the AKLT model \cite{Chandran}. 
In low energy excited eigenstates, such triplon excitations are dominant. 
If we regard the triplon excitation as a quasiparticle, the quasiparticles may be localized under a finite disorder.
It is expected that the localization of the quasiparticles does not affect (interact with) the edge modes, 
and hence, the edge modes are protected by localization of the quasiparticles. 
This means that the SPT order of the groundstate survives even in excited eigenstates. 
Here, under the change of $U$, disorder type, and the target sector of the excited eigenstates, 
it is interesting how the SPT order sustains or not.

In our $H_{XXZ}$ model, we observe a certain typical signal of the protection of the SPT order by localization. 
Using the exact diagonalization \cite{ED1,ED2,ED3,ED4}, 
we numerically calculated the low energy spectrum for $L=18$ system with open boundary, $U=2$, $\delta J_{2}=3$ \cite{spinZ}, 
where edge modes are expected to appears on both sides of the edge. 
In the clean system, the groundstate is fourfold degenerate. 
See Fig.~\ref{Setup} (b)-(c), we plot the lowest to 28-th and the 81-th to 100-th spectrum.
We captured fairly fourfold degeneracy of the spectrum. As observed in the inset figure in Fig.~\ref{Setup} (b), 
the deviation is negligibly small despite the existence of the finite size effect. 
Accordingly, the edge mode degeneracy seems to be protected in low energy excited states. 

\textit{Entanglement spectrum and the degeneracy.---}
We obtained the signal of the protection of the SPT order from the spectrum degeneracy.
To verify this phenomenon in detail, we calculate the entanglement spectrum (ES). 
Consider dividing the system into two equal halves (A and B partial system).
The ES $\{\Lambda^{n}_{\ell}\}$ is defined by $\Lambda^{n}_{\ell} = -2 \log |\lambda^{n}_{\ell}|$, 
each $\lambda^{n}_{\ell}$ is a singular value of the Schmidt decomposition for the $n$-th eigenstate, 
then the reduced density-matrix of $n$-th eigenstate is given by $\rho^{n}_{A}=\sum_{\ell}(\lambda^{n}_{\ell})^2 |\ell\rangle_{A} \langle \ell |_{A}$. 
In what follows, we assume that $\{\Lambda^{n}_{\ell}\}$ is ascending--ordered for $\ell$. 
The degeneracy of the low--lying ES characterizes the SPT order. 
It corresponds to the number of the edge modes \cite{Fidkowski}. 
For an interacting SSH model in the clean limit, 
the ES of the groundstate has been extensively studied to characterize the SPT groundstate by the ES \cite{Sirker,BTYe}. 
In the clean limit, we confirmed that the SPT groundstate in the SSH model of $H_{SSH}$ with $J_2>J_1$ and open boundary 
has two or fourfold degeneracy of the lowest ES depending on the system size $L$.
\begin{figure*}[t]
\centering
\includegraphics[width=17cm]{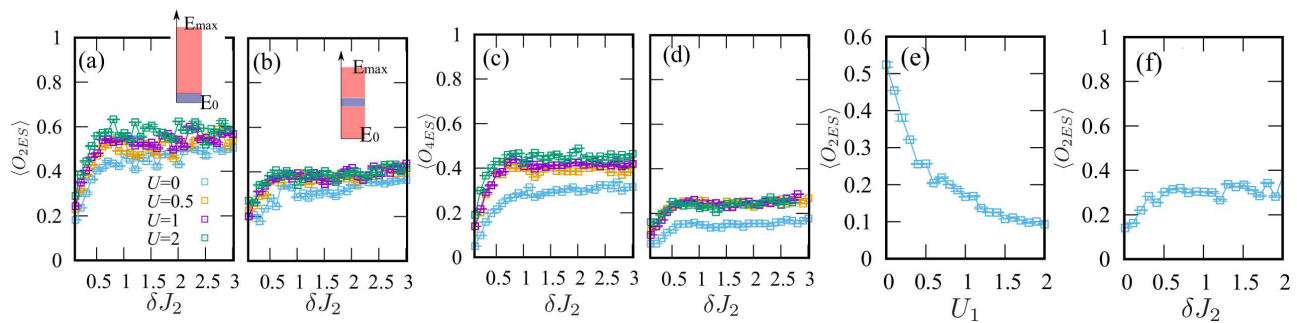}
\caption{The behavior of the ES degenerate order parameters 
with varying disorder strength $\delta J_{2}$ and an interaction form. 
The behavior of $\langle O_{2ES}\rangle$ in $L=14$ system for the low excited energy sector (a) and for the middle excited energy sector (b)
The behavior of $\langle O_{4ES}\rangle$ in $L=12$ system for the low excited energy sector (c) and for the middle excited energy sector (d) 
As a whole, the interaction $U$ enhances the saturation values of $\langle O_{2ES}\rangle$ and $\langle O_{4ES}\rangle$ for large $\delta J_{2}$.
(e) The effects of an odd-link interaction $U_{1}$ for $L=14$ system with $U=2$, $\delta J_{2}=3$ and $\delta J_{1}=0$ for the low excited energy sector. 
(f) Modulated-disordered interaction case. $U_{i}=2J_{i}$. $L=14$ system with $\delta J_{2}=3$ and $\delta J_{1}=0$ for the low excited energy sector. 
For all cases, $10^{2}$ disorder samples were used. Error bars are small and are included in the point labels.}
\label{Fig3}
\end{figure*}

Let us now consider the disordered system. 
We focus on effects of $\delta J^{e}_{i}$ disorder in the low excited energy sector, and
therefore, switch off $\delta J^{o}_{i}$ and $h_i$ disorders.
A periodic boundary condition is considered. 
In the calculation of the ES, we virtually cut the system into two equal halves with the system size $L/2$. 
Here, for $L/2= odd$, it is necessary to cut one $J^{o}_{i}$--link and one $J^{e}_{i}$--link to obtain two equal halves. 
If the system is in the SPT phase, the partial system has a single free spin--1/2 edge mode on either right or left edge sites. 
This leads to twofold degeneracy of the lowest ES \cite{Sirker}. 
Meanwhile, for $L/2= even$, we cut two $J^{e}_{i}$--links to obtain two equal halves. 
In the SPT phase, this leads to two free spin--1/2 edge modes on both right and left edge sites 
corresponding to fourfold degeneracy of the lowest ES \cite{Sirker}. 
For $L=14$ system with $U=0$ and $\delta J_{2}=0$, 
we first calculated the ES by using many-body eigenstates from groundstate to 100-th excited state. 
The lowest and second lowest ES, $\Lambda_{1}$ and $\Lambda_{2}$ are plotted in Fig.~\ref{ES_degenracy} (a). 
While $\Lambda_{1}$ and $\Lambda_{2}$ are degenerate for the groundstate, 
$\Lambda_{1}$ and $\Lambda_{2}$ for many other excited states are not degenerate. 
This indicates that the SPT order vanishes in low energy excited states except for the groundstate. 
In contrast, see an interacting disorder case $\delta J_2=3$ as shown in Fig.~\ref{ES_degenracy} (b). 
This is a single-shot result for a particular $\delta J^{e}_{i}$ disorder distribution.
Interestingly, we observed the proliferation of twofold degeneracy of the ES even in low energy excited eigenstates. 
As shown in inset Fig.~\ref{ES_degenracy} (b), the deviation of the degeneracy is negligibly small even in the finite system size.
The result indicates the survival of the edge modes even in low energy excited states due to $\delta J^{e}_{i}$ disorder. 
Furthermore, we also calculated the ES of $L=16$ system for a noninteracting clean case in Fig.~\ref{ES_degenracy} (c). 
The lowest to fourth lowest ES, $\Lambda_{1}$, $\Lambda_{2}$ $\Lambda_{3}$ and $\Lambda_{4}$ are plotted. 
As observed in Fig.~\ref{ES_degenracy} (a), for the groundstate, the fourfold degeneracy from $\Lambda_{1}$ to $\Lambda_{4}$ appears, 
while for many other excited states the fourfold degeneracy is not observed. 
However, as shown in Fig.~\ref{ES_degenracy} (d), we calculated the ES for an interacting disorder case $\delta J_2=3$ 
and found the proliferation of the fourfold degeneracy of the ES even in low energy excited states. 
This indicates  for $L/2=even$ system the survival of the edge modes even in low energy excited eigenstates. 
Thus, the ES calculation indicates that the $\delta J^{e}_{i}$ disorder protects the SPT order in the low energy excited eigenstates.

Further, we investigate how the low--lying ES degeneracy depends on $\delta J_{2}$ and the regime of excited energy density.
To quantify the degree of the degeneracy of the low--lying ES, we define the degree of the two or fourfold degeneracy of the ground ES, 
for $L/2= odd$ system $\Delta \Lambda_{1-2}(n)\equiv \Lambda_{2}-\Lambda_{1}$, for $L/2= even$ system $\Delta \Lambda_{1-4}(n)\equiv 3\Lambda_{1}-\Lambda_{2}-\Lambda_{3}-\Lambda_{4}$, and then define the following measure:
\begin{eqnarray}
O_{2ES}=\frac{1}{N_{s}}\sum_{n\in {\bf N}_{E}}\theta(\epsilon-\Delta \Lambda_{1-2}(n)),\nonumber\\
O_{4ES}=\frac{1}{N_{s}}\sum_{n\in {\bf N}_{E}}\theta(\epsilon-\Delta \Lambda_{1-4}(n)),
\nonumber
\label{SPT_order}
\end{eqnarray}
where $N_{s}$ is a number of eigenstate in the target regime of the spectrum denoted by ${\bf N}_{E}$, $\theta(X)$ is the Heaviside function, 
and $\epsilon$ represents a practical numerical criteria of the degeneracy of the ES \cite{epsiES}. 
We calculate the disorder--averaged values of $O_{2ES}$ and $O_{4ES}$ denoted by $\langle O_{2ES}\rangle$ and $\langle O_{4ES}\rangle$.
In the disordered system, $\langle O_{2ES}\rangle$ and $\langle O_{4ES}\rangle$ can be regarded 
as an order parameter to characterize the SPT order in the excitation energy parameter space.
Figure~\ref{Fig3} shows various behaviors of $\langle O_{2ES}\rangle$ and $\langle O_{4ES}\rangle$ 
as varying $\delta J_2$ with $\delta J_1=0$.

See Fig.~\ref{Fig3} (a), $\langle O_{2ES}\rangle$ is of $L=14$ system 
and we set ${\bf N}_{E}$ to $10\%$ excited eigenstates from the groundstate in all eigenstates. 
We found that for various $U$, $\langle O_{2ES}\rangle$ increases with the increase of $\delta J_2$, and 
that interestingly $\langle O_{2ES}\rangle$ saturates up to $\sim 0.7$ for large $U$. 
This implies that almost $70\%$ excited eigenstates in the set of ${\bf N_{E}}$ exhibit a single edge mode on the edge site. 
The low energy excited eigenstates are fairly ordered in the SPT order with the increase of $\delta J_2$. 
Furthermore, for large $\delta J_2$, $\langle O_{2ES}\rangle$ of large $U$ case tends to be larger than that of small $U$. 
The result indicates that the interaction $U$ tends to promote the SPT order for excited eigenstates.  
We further calculated $\langle O_{2ES}\rangle$ with the same condition but for the middle $10\%$ excited eigenstates in all eigenstates.
Remarkably, $\langle O_{2ES}\rangle$ also increases as increasing $\delta J_2$, 
the saturate value of $\langle O_{2ES}\rangle$ is somewhat small compared to that in the low energy excited sector.
This indicates that the protection of the SPT order is weak in the middle excited energy sector. 
However, interestingly for the high excited energy sector, 
the protection of the SPT order tends to be strong again, although the interaction dependence varies. 
See the supplemental material \cite{supp}. 
Let us turn on the disorder dependent behavior of $\langle O_{4ES}\rangle$ as shown in Fig.~\ref{Fig3} (c)-(d). 
The results are of $L=12$ system size. The results in Fig.~\ref{Fig3} (c)-(d) have the same tendency as those in Fig.~\ref{Fig3} (a)-(b), and 
therefore, we expect $L/2= even$ system to clearly exhibit the protection of the SPT order in the low of the excited states. 
In addition, we checked the system size dependence for $\langle O_{2ES (4ES)}\rangle$, see \cite{supp}.

Other interaction conditions were also calculated. 
Effect of the odd-link interaction $U_{i=odd}=U_{1}$ is shown in Fig.~\ref{Fig3} (e). 
Here, other parameters are $U=2$ and $\delta J_{2}=3$. 
As approaching a uniform interaction $U_{i}=U$ for all $i$, $\langle O_{2ES}\rangle$ is decreasing. 
The MBL protection is decreasing. The uniform interaction case tends to be close to the claim in \cite{Vasseur}.
Second, $U_{i}=2J_{i}$ case was also calculated as shown in Fig.~\ref{Fig3}. 
The parameters are same with Fig.~\ref{Fig3} (e). 
From the result, as a whole $\langle O_{2ES}\rangle$ is smaller than those of the interacting cases in Fig.~\ref{Fig3} (a) and (c). 
Our numerics shows that the MBL protection is subtle.

\textit{Effects of diagonal disorder and MBL signal.---}
We investigated effects of $\delta h$ disorder. 
This disorder breaks the chiral symmetry in the system of $H_{SSH}$. 
Hence, we expect that this disorder strongly affects the edge mode in the SPT order and that the SPT order tends to vanish with a certain strength of $\delta h$. 
Figure.~\ref{Fig4} (a) shows a typical $\delta h$ dependence of $\langle O_{2ES}\rangle$ for $L=14$ system with $\delta J_2=3$, $\delta J_1=0$. 
For the low and middle sector of the excited energy, $\langle O_{2ES}\rangle$ decreases with the increase of $\delta h$ in various $U$. 
There is no clear spectrum phase transition, but a crossover towards the conventional MBL without the SPT order is observed.
The protection of the SPT is fairly robust, which is almost suppressed at $\delta h\sim 1$, comparable to $J_2$ energy scale. 
We also calculated effects of $\delta J_1$ disorder and obtained a global phase diagram \cite{supp}. 
In addition, we investigated the stability of MBL. 
To characterize the MBL for each excited eigenstates, 
we employ the average level spacing ratio $\langle r \rangle$ \cite{Oganesyan1,Luitz,Janarek}.
Particularly, to obtain the clear behavior of $\langle r\rangle$, 
we focused on the $\sum^{L}_{i=1}S^{z}_{i}=0$ sector in the Hilbert space and on the middle excited eigenstates \cite{Luitz}. 
In the calculation of $\langle r \rangle$, consider the spectrum $\{E_{i} \}$ (in ascending order). 
Each level spacing in $\{E_{i} \}$ is given as $r^{k}=[{\rm min}(\delta^{(k)}, \delta^{(k+1)})]/[{\rm max}(\delta^{(k)},\delta^{(k+1)})]$, 
where $\delta^{(k)}=E_{k+1}-E_{k}$, and then the value of $\langle r \rangle$ is obtained by averaging over disorder samples. 
From the value of $\langle r\rangle$, we can estimate localization in the target sector of the excited eigenstates: 
For the MBL, $\langle r \rangle \sim  0.386$, corresponding to the Poisson random matrix ensemble. 
The results of $\langle r\rangle$ are shown in Fig.~\ref{Fig4} (b)-(c), where we employed $L=16$ system. 
Figure.~\ref{Fig4} (b) is $\delta J_2$-dependence of $\langle r \rangle$ for a finite moderate $\delta h$ case ($\delta h=0.01$, $0.1$ and $0.2$) and $U=2$.  
For small $\delta J_2$, $\langle r\rangle$ takes $\sim 0.53$ the signal of an extended state, corresponding to Gaussian orthogonal ensemble \cite{Luitz,Janarek}, 
and with the increase of $\delta J_2$, the values of $\langle r \rangle$ approach $\sim 0.386$.  
$\delta J_{2}$-disorder induces MBL, i.e., MBL is exhibited when the eigenstates turn into the SPT order as increasing $\delta J_{2}$ from zero \cite{dh=0}.
Figure.~\ref{Fig4} (c) displays $\delta h$-dependence of $\langle r \rangle$ with $\delta J_{2}=3$. 
Even in varying $\delta h$ where the SPT order is suppressed in excited eigenstates, 
the MBL signal $\langle r \rangle \sim 0.386$ remains. 
This indicates that during the crossover behavior in Fig.~\ref{Fig4} (a), the MBL sustains.

\begin{figure}[t]
\centering
\includegraphics[width=8cm]{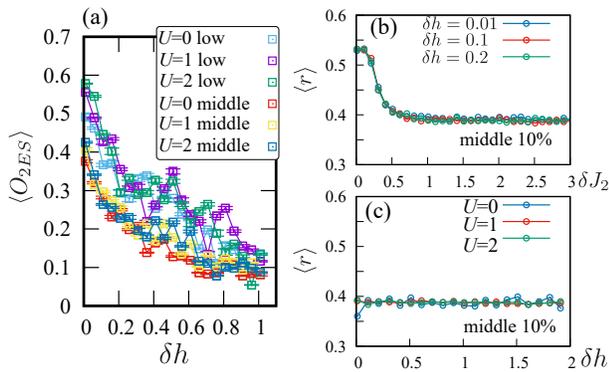}
\caption{(a) $\delta h$ dependence of $\langle O_{2ES}\rangle$ for $L=14$ system 
with $J_1=0.1$ and $J_2=1$. 
Error bars are small and are included in the point labels.
(b) $\delta J_2$-dependence of the average gap ratio $\langle r \rangle$ 
for the middle excited energy sector.
(c) $\delta h$-dependence of the average gap ratio $\langle r \rangle$ 
for the middle excited energy sector. 
For all results, we averaged over $10^2$ disorder samples.
}  
\label{Fig4}
\end{figure}

\textit{Conclusion.---}
For a simple modified XXZ model, we numerically investigated the MBL protection of the SPT phase. 
We found that for a certain interaction pattern, the MBL clearly protects the SPT order in many--body excited eigenstates. 
However, a uniform interaction situation and the diagonal disorder weaken the tendency of the MBL protection. 
Although the limitation in this work is the small system sizes accessible to exact diagonalization, 
our results indicate that the SPT order in the XXZ model possibly appears even in high temperature or out of equilibrium.


\textit{Acknowledgments.---}
Y. K. acknowledges the support of the Grant-in-Aid for JSPS
Fellows (No.17J00486). Y. K. is very grateful to Jakub Zakrzewski for his critical reading of my manuscript and 
for providing numerous helpful suggestions that improved my draft, and thanks an anonymous referee for valuable suggestions.


\section*{\large{Supplemental Material}}
\section{Degeneracy of the ES in high excited eigenstates}
\begin{figure}[h]
\begin{center} 
\includegraphics[width=7.5cm]{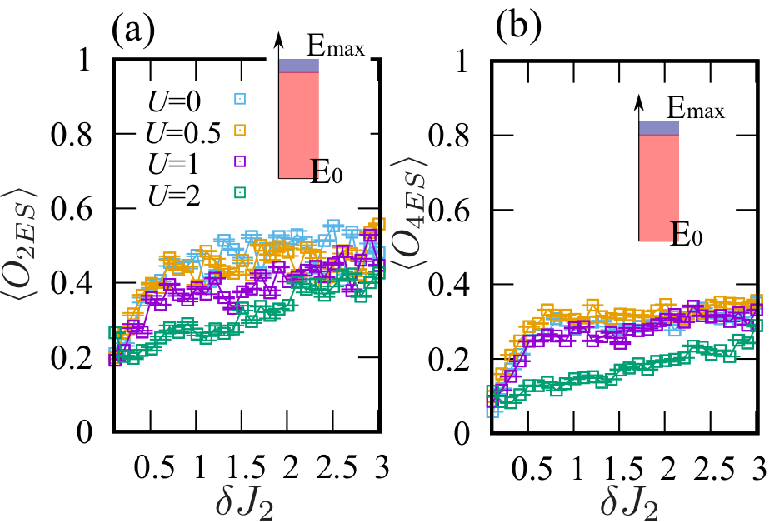}
\end{center} 
\caption{The behavior of the ES degenerate order parameters in varying the disorder strength.
(a) The behavior of $\langle O_{2ES}\rangle$ in $L=14$ system for the high excited energy sector. We used $10^2$ disorder samples. 
(b) The behavior of $\langle O_{4ES}\rangle$ in $L=12$ system for the high excited energy sector. We used $10^2$ disorder samples. 
For the calculation of $\langle O_{2ES}\rangle$, $\epsilon=0.1$ and for the calculation of $\langle O_{4ES}\rangle$, $\epsilon=0.2$.
Error bars are small and are included in the point labels.}
\label{supv2}
\end{figure}

In the main text, we observed the MBL induced protection of the edge mode in the low and middle excited energy sector. 
Here, let us observe the protection behavior for in the high excited energy sector. 
Figure~\ref{supv2} shows various behaviors of $\langle O_{2ES}\rangle$ and $\langle O_{4ES}\rangle$ 
as varying $\delta J_2$ with $J_1=0.1$, $J_{2}=1$, $\delta h=0.01$, and $\delta J_1=0$.
We set ${\bf N}_{E}$ to $10\%$ excited eigenstates below the highest eigenstate in all eigenstates. 
See Fig.~\ref{supv2} (a), $\langle O_{2ES}\rangle$ is of $L=14$ system with $J_1=0.1$, $J_2=1$.
As same with the low and middle excited energy sector, we found that for various $U$,  $\langle O_{2ES}\rangle$ increases with the increase of $\delta J_2$.
However, compared with Fig.~3 (a)-(b) in the main text, the interaction $U$ does not enhance $\langle O_{2ES}\rangle$ for large $\delta J_2$. 
This tendency is also true for the behavior of $\langle O_{4ES}\rangle$ with $L=12$ system as shown in Fig.~\ref{supv2} (b). 
As a whole, the protection of the SPT in the high excited energy sector is stronger than that of the middle excited energy sectors. 
Therefore, the MBL protection tends to be somewhat strong at band edges compared to the center of the band.

\section{Finite system size effects in the behavior of the entanglement spectrum degeneracy}

\begin{figure}[h]
\begin{center} 
\includegraphics[width=6.5cm]{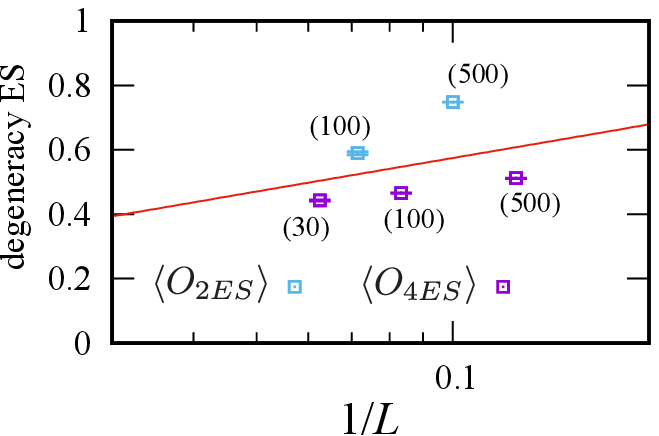}
\end{center} 
\caption{System size dependence of the degeneracy of the ES.  
The number nearby the labels represents the number of the disorder samples. The red line is a fitting line determined by all data.
The red line is $0.1502\ln (1/L) + 0.9201$
}  
\label{sup3}
\end{figure}

We show a typical system size dependence of the degeneracy of the ES. 
We calculated $\langle O_{2ES}\rangle$ and $\langle O_{4ES}\rangle$ for various system size. 
The definition of $\langle O_{2ES}\rangle$ and $\langle O_{4ES}\rangle$ is shown in the main text.
We set the parameters $J_{1}=0.1$, $J_{2}=1$, $U=2$, $\delta J_{1}=0$, $\delta J_{2}=3$ and $\delta h=0.01$, 
where the system exhibits the MBL protection for the SPT phase 
and we used the lowest $10\%$ many-body eigenstates.
The result is shown in Fig.~\ref{sup3}. Here, $L=$8,10,12,14 and 16 systems were employed. 
When the system size is increased, the decrease in the value of $\langle O_{2ES}\rangle$ and $\langle O_{4ES}\rangle$ seems small. 
This fact indicates that the ES degeneracy characterized by a finite value of $\langle O_{2ES}\rangle$ and $\langle O_{4ES}\rangle$ survives for fairly large system size.

\section{$\delta J_2$ and $\delta h$ disorder phase diagram}
\begin{figure}[h]
\begin{center} 
\includegraphics[width=5.5cm]{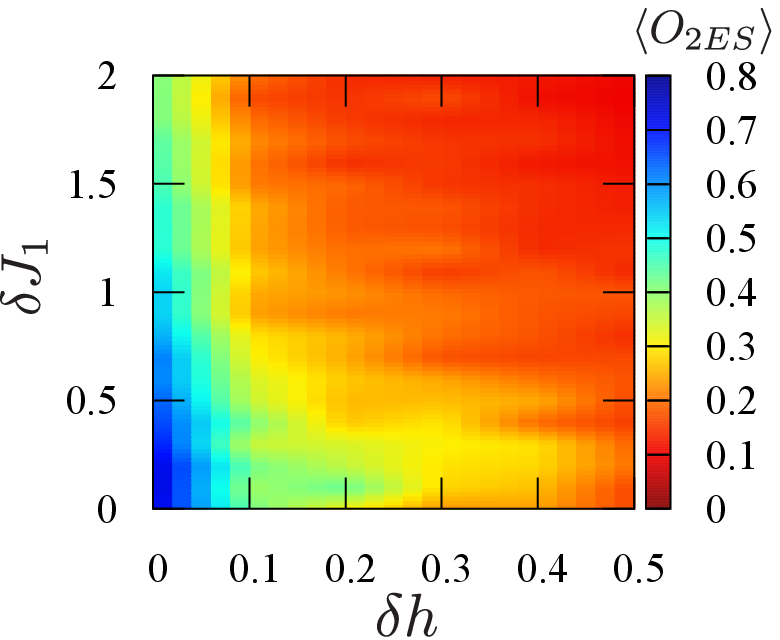}
\end{center} 
\caption{$\delta J_2$ and $\delta h$ disorder phase diagram. 
We used $10^2$ disordered samples.}  
\label{sup2}
\end{figure}

In the main text, we observed the $\delta h$ disorder dependence of $\langle O_{2ES}\rangle$ with $\delta J_2=3$. 
Here we observe effects of $\delta J_1$ disorder by calculating a global phase diagram. 
We employed $L=10$ system with $J_1=0.1$, $J_2=1$, $U=2$, and $\delta J_2=3$, 
set ${\bf N}_{E}$ to $10\%$ excited eigenstates from the groudstate in all eigenstates, 
and calculated $\langle O_{2ES}\rangle$ by averaging over disorder samples.
The result is shown in Fig.~\ref{sup2}. 
For small $\delta h$, the $\delta J_{1}$ disorder does not suppress the value of  $\langle O_{2ES}\rangle$ so much.
In contrast, the $\delta h$ disorder more suppresses the value of  $\langle O_{2ES}\rangle$. 
Therefore, the MBL protection of the SPT order are fairly robust to effect of the $\delta J_{1}$ disorder.
It is natural because the $\delta J_{1}$ disorder does not break the chiral symmetry of the system of $H_{XXZ}$ 
and therefore the SPT order is hardly affected.

\begin{thebibliography}{99}
\bibitem{edge_mode}
In the thermodynamic limit, the edge mode exhibits exact zero--energy excitation. 
For a finite size case, the energy--excitation is exponentially small with the system size $L$.
\bibitem{degeneE}
In our numerics, if we set $\delta h=0$, degeneracy in the spectrum is proliferated. 
This may originate from the symmetries of the topological model 
and the non-conservation of the total spin-z component in the system. 
Such a degeneracy can lead difficulty in evaluating the degeneracy of the entanglement spectrum. 
Once we inject small but finite $\delta h$, the degree of the degeneracy is expected to be lessened.
\bibitem{spinZ}
In all numerical simulation except for Fig.~\ref{Fig4} (b)-(c), the Hilbert space includes all spin z-component sectors. 
The total spin z-component $S^{z}_{total}=\sum_{i}S^{z}_{i}$ is not conserved.
\bibitem{epsiES}
As shown in the inset figures of Fig.~\ref{ES_degenracy} (b) and (d), since the deviation of the degeneracy coming from finite size effects is negligibly small, 
we set $\epsilon=0.1$ for the calculation of $\langle O_{2ES}\rangle$ and  $\epsilon=0.2$ for the calculation of $\langle O_{4ES}\rangle$.
\bibitem{dh=0}
For $\delta h=0$ case,  $\langle r \rangle$ indicated less than $ 0.386$.
This comes from the fact that the level spacing ratio is difficult to capture the signal of the MBL 
when the high degeneracy of the entire spectrum of the model of $H_{XXZ}$ without $h_{i}$ disorder remains.
However, we expect since very small $\delta h=0.01$ case clearly indicates the MBL signal and therefore even for the limit $\delta h\to0$, 
the system with only $\delta J_2$ disorder is in the MBL.

\clearpage

\renewcommand{\thesection}{S\arabic{section}} 
\renewcommand{\theequation}{S\arabic{equation}}
\renewcommand{\thefigure}{S\arabic{figure}}
\setcounter{equation}{0}
\setcounter{figure}{0}
\end{thebibliography}
\end{document}